\documentclass[useAMS,usenatbib]{mn2e}
\usepackage[pdftex]{graphicx} 
\usepackage[hidelinks]{hyperref}
\usepackage{xcolor} 
\usepackage[T1]{fontenc}
\usepackage{aecompl}

\hypersetup{colorlinks,linkcolor={red!50!black}, citecolor={blue!50!black},urlcolor={blue!80!black} } 

\title[Testing galaxy formation models]
{Testing galaxy formation models with galaxy stellar mass functions}

\author[S.H. Lim et al.]{S.H. Lim$^{1}$\thanks{E-mail:
slim@astro.umass.edu}, H.J. Mo$^{1,2}$, Ting-Wen Lan$^{3,4}$ and Brice M\'enard$^{3,4}$\\
$^{1}$Department of Astronomy, University of Massachusetts, Amherst MA 01003-9305, USA \\
$^{2}$Physics Department and Center for Astrophysics, Tsinghua University, Beijing 10084, China\\
$^{3}$Kavli IPMU (WPI), UTIAS, The University of Tokyo, Kashiwa, Chiba 277-8583, Japan \\
$^{4}$Department of Physics \& Astronomy, Johns Hopkins University, Baltimore MD 21218, USA}

\begin{document} 

\date{Accepted ........ Received .......; in original form ......}

\pagerange{\pageref{firstpage}--\pageref{lastpage}}

\pubyear{2016}

\maketitle

\label{firstpage}

\begin{abstract} 
We compare predictions of a number of empirical models and numerical 
simulations of galaxy formation to the conditional stellar mass functions (CSMF)
of galaxies in groups of different masses obtained recently by Lan et al. 
to test how well different models accommodate the data. 
The observational data clearly prefer a model in which star formation in low-mass halos 
changes behavior at a characteristic redshift $z_c\sim 2$. There is also 
tentative evidence that this characteristic redshift depends on environment, 
becoming $z_c\sim 4$ in regions that eventually evolve into rich clusters 
of galaxies. The constrained 
model is used to understand how galaxies form and evolve in 
dark matter halos, and to make predictions for other statistical properties 
of the galaxy population, such as the stellar mass functions of galaxies 
at high $z$, the star formation and stellar mass assembly histories in 
dark matter halos. A comparison of our model predictions with those 
of other empirical models shows that different models can make vastly different 
predictions, even though all of them are tuned to match the observed 
stellar mass functions of galaxies. 

\end{abstract} 

\begin{keywords} 
methods: statistical -- galaxies: formation -- galaxies: evolution -- galaxies: haloes.
\end{keywords}

\section{Introduction}
\label{sec_intro}

In the current paradigm of structure formation within the $\Lambda$ cold dark matter 
($\Lambda$CDM) framework, initial small fluctuations in the cosmic density field are 
amplified by gravitational instability, eventually forming highly nonlinear structures
called dark matter halos (see \citet{mo10} for a review). Galaxies then form 
at the centers of the gravitational potential wells of the dark matter halos by 
radiative cooling and condensations of baryonic gas \citep[e.g.][]{white78, fall80, 
mo98}.  In order to reproduce the observed 
stellar mass function of galaxies in the CDM scenario, however, star formation in 
dark matter halos has to be inefficient \citep[e.g.][]{yang03}, 
and various feedback processes have been proposed to suppress the star formation 
efficiency in dark matter halos.  

In this framework, therefore, galaxy formation and evolution are governed 
by a number of physical processes which, in turn, are characterized by a number 
of characteristic scales. First, cosmological $N$-body simulations have 
shown that the assembly histories of dark matter halos in general consist of 
two distinctive phases: an earlier phase of fast mass acquisition 
during which the potential well of a halo deepens rapidly with time, and a 
later phase of slow accretion, with a time scale longer than the Hubble 
time \citep[e.g.][]{zhao03}. \citet{zhao09} found that the two phases 
are separated at a time when a halo obtains about $\sim 4\%$ of its final 
mass \citep[see also][]{vdB14}. Second, hydrodynamical simulations 
have demonstrated that radiative cooling is effective in halos with masses 
smaller than $M_h\sim 6\times 10^{11} {\rm M_\odot}$, so that the accretion rate 
of cold gas into galaxies is determined by the halo mass accretion rate, 
independent of radiative cooing \citep[e.g.][]{keres05, keres09}. 
Above this mass scale, on the other hand, radiative cooling is ineffective, so that 
the cold gas accretion is delayed by the cooling time scale. For massive halos 
with masses above $10^{13}{\rm M_\odot}$, a significant fraction of the baryonic gas is expected 
to be in the hot halo in the absence of a heating source. Third, supernova feedback from star formation 
is believed to be effective for halos with masses below $\sim 10^{11}{\rm M}_\odot$
\citep[e.g.][]{dekel86, somerville08, lu12}. Finally 
AGN feedback from accreting super-massive black holes has been proposed as 
a mechanism to suppress star formation in massive halos, with masses above 
$M_h\sim 10^{13} {\rm M_\odot}$ \citep[e.g.][]{ferrarese00, mcconnell11}. 

A number of approaches have been adopted to explore the physical processes 
that govern galaxy formation and evolution, and to facilitate 
comparisons between theory and observation.
The first is hydrodynamical simulation that includes
both dark matter and baryonic components \citep[e.g.][]{dubois14, 
khandai15, vogelsberger14, schaye15}. However, due to limited resolution and subgrid 
implementations of some key processes, the results obtained from such simulations 
are still questionable, even though they can match some observational data 
\citep[e.g.][]{governato04, governato10, okamoto05, guedes11}. 
Furthermore, high resolution hydrodynamical simulations are computationally 
expensive, which prohibits the explorations of a large parameter space.   
Because of this, an alternative approach, the semi-analytic model
(SAM) of galaxy formation, has been developed \citep[e.g.][]{white91, 
kauffmann99, kang05, bower06, croton06, somerville08, guo11, lu11}. 
The SAM approach combines halo merger histories, obtained either 
from dark-matter only simulations or from analytical models, with gas and 
star formation processes using parametrized functions that describe the underlying 
physical processes. This approach is computationally inexpensive, allowing one 
to investigate a large set of different models. However, since all the 
physical processes are approximated with simple empirical functions, 
the reliability and accuracy of this approach needs to be checked.  More recently,
a third approach has been adopted to understand how galaxies form and evolve 
in the cosmic density field. The goal of this approach is to establish the 
connections between galaxies and dark matter halos through an empirical approach, 
using observational data as constraints. 
Models developed along this line include the halo occupation distribution 
\citep[HOD; e.g.][]{jing98, peacock00, white01, berlind02, bullock02, 
zehavi04, zehavi11}, 
the conditional luminosity function \citep[CLF;][]{yang03, yang12, vdB03}, 
the halo abundance matching model \citep[HAM;][]{kravtsov04, vale04, vale06,
conroy06, behroozi10, guo10, moster10, reddick13}, 
and the halo-based empirical model \citep[][]{lu14,lu15}.

To a certain degree, both the SAM and empirical approaches are methods to summarize 
observational data in terms of model parameters characterizing the galaxy-halo
connections. Much progress has been made recently in this area. 
Using the CLF model and constraints of the observed luminosity function and 
correlation function of galaxies, \citet{yang03} found a characteristic 
halo mass scale, $\sim 10^{12}{\rm M_\odot}$, in the relationship between galaxy 
luminosity/stellar mass and halo mass relation, suggesting that star formation 
efficiency declines rapidly toward both the higher and lower mass ends. 
With the use of galaxy groups selected from the 2dF \citep[][]{yang05}
and SDSS \citep[][]{yang07}, \citet{yang05} found a similar mass 
scale from the observed galaxy luminosity/stellar mass - 
halo mass relations obtained directly from galaxy groups.  
In particular, \citet{yang05} suggested the existence of 
another characteristic mass scale, 
$\sim 10^{11}{\rm M_\odot}$, where the galaxy luminosity-halo 
mass relation may change its behavior. Similar results 
have since been obtained at higher $z$ with the use of the 
observed luminosity/stellar mass functions of galaxies. 
In particular, the presence of the mass scale at $\sim 10^{12} {\rm M_\odot}$ 
seems to extend to higher $z$ without showing 
strong evolution \citep[][]{moster10, behroozi13, guo10, yang12}.  

More recently, \citet{lu14, lu15} developed a halo-based empirical 
model to follow the star formation and stellar mass assembly histories of 
galaxies in dark matter halos. In particular, they used the observed 
conditional luminosity functions of cluster galaxies obtained by 
\citet{popesso06} as an constraint in addition to the field stellar mass functions 
at different redshifts. They found that the observational data require 
two additional characteristic scales, a characteristic redshift, $z\sim 2-3$,  
and a corresponding mass scale at $10^{11}{\rm M_\odot}$, below which star 
formation changes behavior at the characteristic redshift. These results 
clearly demonstrate that the observed conditional luminosity/stellar mass function 
of galaxies in clusters can provide important information about galaxy 
formation and evolution at high redshift. However, since clusters of 
galaxies only contain a small fraction of the total galaxy population, 
the results may be affected by some environmental effects that are 
specific only to clusters of galaxies. 

Using the galaxy groups of \citet{yang07} combined with galaxies in the SDSS 
photometric catalogue, \citet{lan16} have recently measured 
the conditional luminosity/stellar mass functions  
(hereafter CSMFs) that cover four orders of magnitude in galaxy luminosity, 
and three orders of magnitude in halo mass, from $\sim 10^{12}$ to 
$10^{15} {\rm M}_\odot$. They found a characteristic luminosity scale, 
$L \sim 10^9 L_\odot$, below which the slope of the CSMF
becomes systematically steeper, and that this trend is present for all halo masses. 
This ubiquitous faint-end upturn suggests that it is formation, rather than 
cluster-specific environmental effect, that plays the dominating 
role in regulating the stellar masses of faint satellites. Clearly, 
these observational results will provide new constraints on models. 

This paper consists of two parts. First, we use the new CSMFs to 
update the empirical model of \citet{lu14, lu15} and show that 
there is only marginal difference between the original model and the 
updated model. Second, we compare model predictions from empirical
models and numerical simulations to the CSMFs of Lan et al. to test 
how well different models accommodate the new data. We will show that, 
among all the models considered, only the \citet{lu14, lu15} model 
can match the observational data reasonably well. Also, we present
predictions of the different models for other statistical properties of the galaxy 
population.

The organization of this paper is as follows. 
Section \ref{sec_model} describes the empirical models to be tested and 
two recent numerical simulations, Illustris \citep[][]{vogelsberger14}
and EAGLE \citep[][]{schaye15}, to be compared. 
In Section \ref{sec_data}, we describe the observational data that are 
used in our analysis, and present comparisons of the empirical models 
and the simulations with them. 
In Section \ref{sec_impl}, we present a more 
detailed comparison of the model predictions in star formation rate, 
stellar mass - halo mass relation, mass assembly history, and stellar mass function 
for high redshifts. Finally, we summarize and discuss our results in 
Section \ref{sec_sum}.

\section[model]{Models}
\label{sec_model}

In this paper we select a number of popular empirical models and two recent 
hydrodynamical simulations to test against observational data. 
Here we describe these models and simulations briefly. 
Table 1 lists the models and the simulations that we test.
Readers are referred to the original papers for details.   

\begin{table}
 \renewcommand{\arraystretch}{1.5} 
 \centering
 \begin{minipage}{65mm}
  \caption{A list of the models and the simulations.}
  \begin{tabular}{ccc}
\hline
Model & Reference  \\
/ Simulation &  \\
\hline
\hline
Y12 & \citet{yang12} \\ 
M13 & \citet{moster13} \\
B13 & \citet{behroozi13} \\
L15 & \citet{lu15} \\
L15-U & this work \\
Illustris & \citet{vogelsberger14} \\
EAGLE & \citet{schaye15} \\

\hline
\\
\vspace{-10mm}
\end{tabular}
\footnotesize{}
\end{minipage}
\label{tab_model}
\end{table}

\subsection{Empirical models} 

One of the simplest way to link galaxies to their dark matter halo/subhalo 
population is to use halo abundance matching \citep[e.g.][]{mo99}. 
This approach assumes
a monotonic relation between halo mass and galaxy stellar mass. 
Satellite galaxies observed at a given redshift were all once central galaxies 
before they were accreted onto larger halos. Since satellites are expected 
to evolve differently due to environmental effects such as tidal stripping and 
ram pressure stripping, many abundance matching models
apply a monotonic relation between galaxy stellar mass 
and halo mass at the time when a halo first became a subhalo, 
instead of at the time of observation. Most of previous investigations 
make the assumption that the halo mass - galaxy mass relation is 
independent of when a sub-halo is accreted into its host
\citep[e.g.][]{vale04, vale06, conroy06, behroozi10, guo10, moster10}. 
With this assumption, at a given redshift, halos of a given mass are 
therefore always linked to galaxies of the same stellar mass. 
However, it was found that applying this method to different redshifts 
actually leads to different stellar mass - halo mass relation \citep[e.g.][]{conroy06}, 
suggesting that the method implemented in this way is not self-consistent. 
As an improvement, models have been developed in which the galaxy-halo 
relation is allowed to depend on both halo mass (defined e.g. 
at the time when a halo first becomes a sub-halo) and the time 
when a halo becomes a sub-halo. 
We test four models in this category, by \citet{yang12}, 
\citet{moster13}, \citet{behroozi13}, and \citet{lu15}, respectively.

\subsubsection{Yang et al. model}

The model of \citet{yang12} (Y12, hereafter) takes the same 
functional form as that proposed in \citet{yang03} for the 
halo mass - galaxy luminosity/stellar mass relation:
\begin{eqnarray}\label{eq_mhvsms}
\frac{M_*}{M_h} = N\left[\left(\frac{M_h}{M_1}\right)^{-\beta}+
\left(\frac{M_h}{M_1}\right)^{\gamma}\right]^{-1}. 
\end{eqnarray}
This is basically a double power law specified by two asymptotic slopes, 
$\beta$ and $\gamma$, describing the low- and high-mass end behaviors, respectively, 
and by a characteristic mass scale $M_1$ where the transition between the
two power laws occurs, and with $N$ being an overall amplitude.
The four free parameters were assumed to be redshift dependent and the dependencies 
were modeled by simple functions. The above relation was used to 
assign stellar masses to halos at different redshifts. They adopted 
the halo mass function of \citet{sheth01} to model the halo population. 
For sub-halos, the model of \citet{yang11} was used to follow 
both the mass function and the distribution in the accretion time 
(the time when a halo first becomes a subhalo). A stellar mass is assigned 
to a sub-halo at the time of accretion according to its mass at that time
using equation (\ref{eq_mhvsms}). The subsequent evolution of the satellite 
associated with a sub-halo was followed according to its orbit determined through 
a dynamical friction model. The model parameters were then obtained by fitting the model 
predictions to the observed stellar mass functions (SMFs) of galaxies from 
$z=0$ to $4$, and the correlation function of $z\sim 0$ galaxies as a function of galaxy 
luminosity/stellar mass. 

\subsubsection{Moster et al. model}

\citet{moster13} (M13) adopted a similar double power-law for the 
stellar mass - halo mass relation as described by equation (\ref{eq_mhvsms}), 
and simple functional forms to describe the redshift dependencies of the
model parameters. They applied the relation to halos and sub-halos 
obtained from $N$-body simulations. Individual halos and sub-halos are matched 
and traced across different snapshots (i.e. different redshifts), 
so that merger trees are generated to track their evolutions. 
Galaxies hosted at the centers of halos and sub-halos were referred 
to as centrals and satellites, respectively. For centrals, the stellar masses
were given by the stellar mass - halo mass relation using the 
redshift and halo mass at the snapshot in question. For satellites,
the stellar masses were obtained by applying the stellar mass - halo mass 
relation at the redshift when their halos first became sub-halos 
using their halo masses at this redshift, as in Y12. 
The stellar mass of a satellite
was assumed to remain unchanged in the subsequent evolution.
Some uncertainties in the stellar mass - halo mass relation were taken 
into account. Model parameters characterizing the stellar mass - halo 
mass relation were then tuned to match a set of observed SMFs from $z=0$ to $4$.

\subsubsection{Berhoozi et al. model}

The approach adopted by \citet{behroozi13} (B13) was similar to those 
of Y12 and M13, but the stellar mass - halo mass relation assumed was more complicated
and was designed in part to reproduce the observed SMFs at the faint ends.  
Here again, halo merger trees extracted from $N$-body simulations
were used to trace the formation of dark matter halos. As in Y12 and M13, 
they applied their stellar mass - halo mass relation to `infall' mass at the 
time of accretion to assign stellar masses to subhalos. 
Subsequent stellar mass loss of satellites after their accretion
into their host halos was also taken into account. Finally, they used the observed SMFs 
at $z=0$ - $8$, as well as the cosmic star formation rates and specific 
star formation rates, to constrain their model parameters. 

\subsubsection{Lu et al. model} 

 The \citet{lu14, lu15} model (hereafter L15) was based on the star formation rate (SFR)
- halo mass relation as a function of redshift: 
\begin{eqnarray}\label{eq_Lu1}
\dot{M_*}(M_h,z) &=& \varepsilon \frac{f_bM_h}{\tau}(x+1)^\alpha 
\Big(\frac{x+R}{x+1}\Big)^\beta\Big(\frac{x}{x+R}\Big)^\gamma
\end{eqnarray}
where $f_b=\Omega_{b,0}/\Omega_{m,0}$, $\tau=[10H_0(1+z)^{3/2}]^{-1}$ approximates the  
dynamical time of halos, $x\equiv M_h/M_c$, with $M_c$ being 
a characteristic mass scale and $R$ is parameter of $0\leq R\leq 1$. 
Thus, $\dot{M_*}/M_h\propto M_h^{\{\alpha, \beta, \gamma\}}$ 
for \{$M_h\gg M_c$, $RM_c<M_h<M_c$, and $M_h\ll RM_c$\}, 
respectively. This relation is applied only to central galaxies. 
After a galaxy becomes a satellite, Lu et al. assumed that it moves 
on an orbit determined by its initial energy and 
orbital angular momentum together with dynamical friction.
A satellite galaxy is assumed to merge with the central galaxy once it 
sinks to the center of the halo. At this time, it adds a fraction 
(treated as a free parameter, $f_{sc}$) of its mass to the central galaxy, and 
the rest is assumed to become halo stars. The SFRs in satellites were
modeled with a simple exponential model, 
\begin{eqnarray}\label{eq_Lu3}
\dot{M}_{*,sat} \propto \exp\Big[-\frac{t-t_{acc}}{\tau_s}\Big]
\end{eqnarray}
where $t_{acc}$ is the time when the galaxy becomes a satellite, 
and $\tau_{s}=\tau_{s,0}\exp\big[-{M_*}/{M_{*,c}}\big]$ 
is adopted to reflect halo mass dependence of the time scale, 
with $\tau_{s,0}$ and $M_{*,c}$ being free parameters. 
The stellar mass in a galaxy is then obtained by integrating the 
SFR over time, taking into account mass loss due to stellar 
evolution. Lu et al. used halo merger trees generated with the algorithm
developed by \citet{parkinson08}, which is based on the extended
Press-Schechter formalism calibrated with $N$-body simulations.

Lu et al. adjusted both their functional forms and free parameters to match
the SMFs at $0<z<4$ and the CSMFs of 
galaxies in clusters of galaxies as given by \citet{popesso06}. 
They found that the model assuming all the parameters to be independent 
of redshift is not able to match the observed SMFs at high redshift. They
therefore extended their model by allowing $\alpha$ to change with redshift
as $\alpha=\alpha_0(1+z)^{\alpha^\prime}$. This model was referred to as 
Model II in Lu et al..  Model II was found to be able to describe all the stellar mass functions (SMFs) 
at both low and high redshifts, but fails to match the faint-end 
upturn in the CSMF of cluster galaxies. Because of this, Lu et al. extended 
their model once more by allowing the parameter $\gamma$, which  
dictates the SFR in low-mass halos, to depend on redshift:  
\begin{eqnarray}\label{eq_Lu4}
\gamma &= &\gamma_a \hspace{13em}{\rm if}\ z\leq z_c \nonumber \\
              &= &(\gamma_a-\gamma_b)\Big(\frac{1+z}{1+z_c}\Big)^{\gamma^\prime} \hspace{5.1em}{\rm if}\ z>z_c \nonumber
\end{eqnarray}
In this model, referred to as Model III by Lu et al., 
$\gamma\to\gamma_b$ at $z\gg z_c$, and the free parameter, 
$\gamma^\prime$, controls how rapidly the transition to $\gamma_b$ occurs 
above the characteristic redshift $z_c$. This Model III was found 
to be able to fit both the SMFs at different redshifts and the CSMF 
of cluster galaxies simultaneously. 

\subsubsection{Updating the parameters of the L15 model}

Instead of using the model parameters of Lu et al., we use only the observed CSMFs
as constraints to update the model parameters. We use the MULTINEST method developed by 
\citet{feroz09}, which makes use of the nested sampling algorithm of 
\citet{skilling06}, to compute the posterior distribution of the model parameters.  
The MULTINEST is found to yield practically the same results as the traditional  
MCMC method but with $\sim10$ times smaller number of likelihood calculations
for the problem concerned here. The reader is referred to the original papers for 
details. 

\begin{figure}
\includegraphics[width=0.94\linewidth]{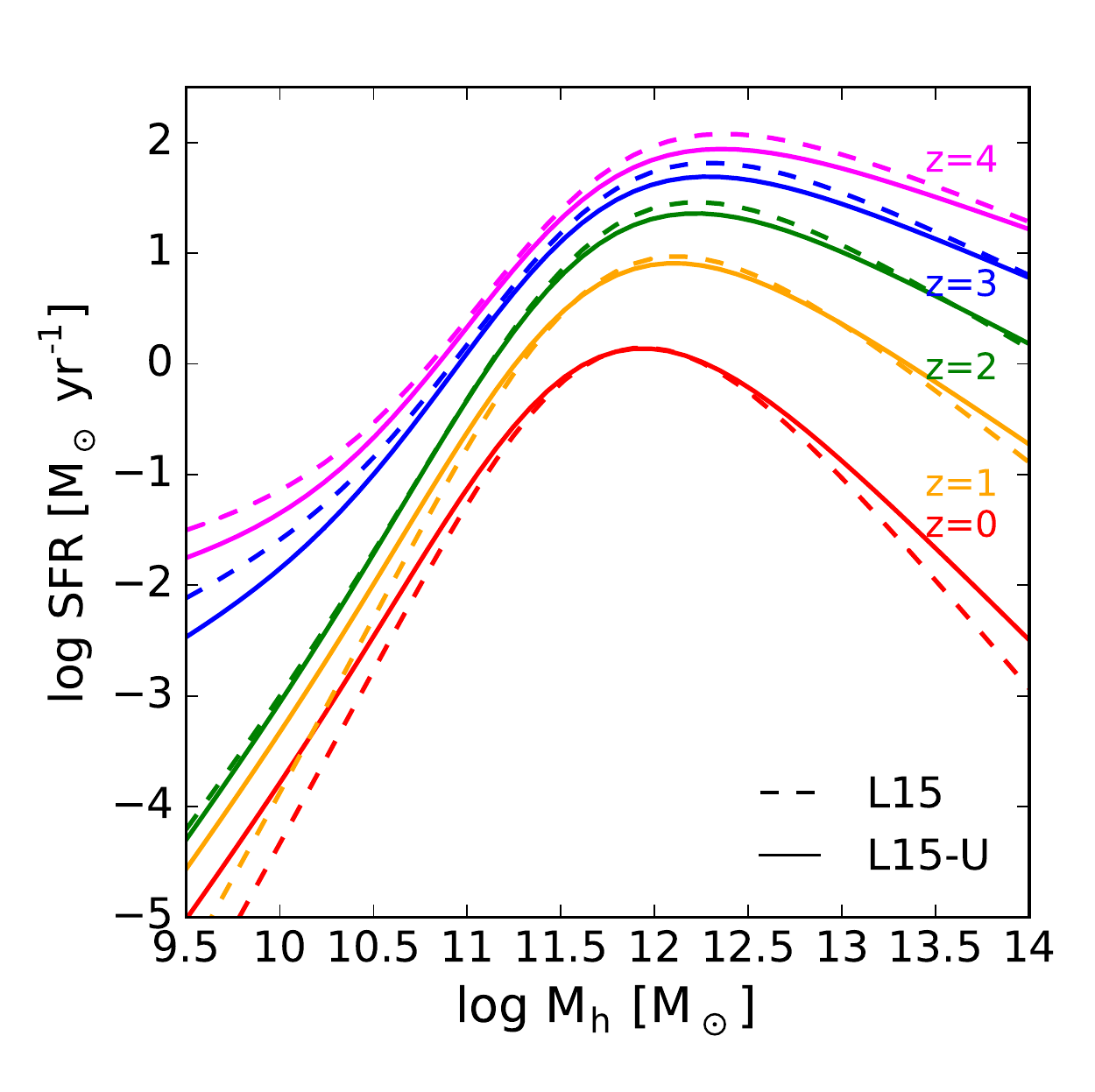}
\caption{The average star formation rate of central 
galaxies as a function of halo mass at different 
redshifts as predicted by the original L15 model (dashed lines) and
the L15-U model (solid lines).}
\label{fig_SFR_L15}
\end{figure}

\begin{table}
 \renewcommand{\arraystretch}{1.5} 
 \centering
 \begin{minipage}{65mm}
  \caption{A list of the model parameters. The medians and the standard deviations
  are presented. }
  \begin{tabular}{ccc}
\hline
parameters & L15 & L15-U \\
\hline
\hline
$\alpha_0$ & $-3.0\pm 1.0$ & $-2.7\pm 0.79$ \\

$\alpha^\prime$ & $-0.36\pm 0.16$ & $-0.32\pm 0.21$ \\

$\beta$ & $3.7\pm 0.73$ & $3.5\pm 1.0$ \\

$\gamma_a$ & $2.0\pm 0.55$ & $1.3\pm 0.69$ \\

$\gamma_b$ & $-0.84\pm 0.14$ & $-1.1\pm 0.21$ \\

$\gamma^\prime$ & $-4.4\pm 0.52$ & $-3.1\pm 0.88$ \\

$z_c$ & $1.8\pm 0.31$ & $2.0\pm 0.28$  \\

$\log_{10}M_c$ & $1.6\pm 0.15$ & $1.6\pm 0.11$ \\

$\log_{10}R$ & $-0.86\pm 0.18$ & $-0.92\pm 0.20$ \\

$\varepsilon$ & $0.20\pm 0.29$ & $0.050\pm 0.11$ \\

$\log_{10}H_0\tau_{s,0}$ & $-0.90\pm 0.16$ & $-0.85\pm 0.11$ \\

$\log_{10}M_{*,c}$ & $0.34\pm 0.28$ & $0.18\pm 0.19$ \\

$f_{sc}$ & $0.44\pm 0.22$ & $0.52\pm 0.15$ \\

\hline
\\
\vspace{-5mm}
\end{tabular}
\footnotesize{}
\end{minipage}
\label{tab_param}
\end{table}

Table 2 compares the updated parameters with the model 
parameters of Lu et al. The average star formation rates predicted with the updated 
parameters at various redshifts are very similar to those obtained by L15, as shown in 
Figure~\ref{fig_SFR_L15}. We also found that the differences in the two 
parameter sets result only in marginal changes in the CSMFs in that the updated 
model (hereafter L15-U) predicts slightly flatter slopes at the faint-ends for massive haloes. 
This is owing to the fact that the Lan et al. CLFs have shallower faint-end slopes 
for massive halos than the cluster galaxy luminosity function used by L15.
The marginal difference between the two parameter sets demonstrates that the 
low-$z$ CSMFs alone can constrain models in a similar way as the field SMFs 
at different redshifts. Furthermore, as we will see in \S\ref{sec_impl}), 
they also contain information about the low-mass end of the SMF at high $z$, 
where direct observations are still uncertain. We use L15-U to present
results throughout this paper.

\subsubsection{Need for a more extended model family?}

As mentioned above, L15 assumed the characteristic redshift, 
$z_c$, the redshift at which the SFR in low mass progenitors changes behavior, 
to be independent of the host halo mass. However, it is plausible that $z_c$ 
depends on the host halo mass, because structure formation, and presumably 
star formation, are expected to occur earlier in regions that correspond to 
higher mass halos at the present day. Motivated by this, we test a more 
extended model family in which the characteristic redshift changes with host halo mass at $z=0$, $M_h(0)$:
\begin{eqnarray}\label{eq_zeta}
(1+z_c) = (1+z_{c,0})\Big(\frac{M_h(0)}{10^{12}{\rmn M_\odot}}\Big)^\zeta 
\end{eqnarray}
where $\zeta$ controls the halo mass dependence of $z_c$, 
and $z_{c,0}$ is $z_c$ for halos of $M_h(0)=10^{12}{\rmn M_\odot}$. 
We use the same CSMFs as used in the earlier subsection to constrain model
parameters.

To test if such an extension is necessary, we use the Bayes factor, 
\begin{eqnarray}\label{eq_bayes}
K = \frac{P(D|M_1)}{P(D|M_2)} = \frac{\int{P(D|\theta_1,M_1)P(\theta_1|M_1)d\theta_1}}
{\int{P(D|\theta_2,M_2)P(\theta_2|M_2)d\theta_2}}, 
\end{eqnarray}
where $D$ is a given data set, $M_1$ and $M_2$ are two different models, and 
$\theta_1$ and $\theta_2$ are the parameter space of the models. 
This factor quantifies the preference of a given data set for 
one model family over the other. As it integrates over all parameter 
space of each of the model families, it naturally penalizes over-fitting. 

When all the data points of the CSMFs are used as constraints, the Bayes factor between the 
extended model (the one including $\zeta$) and the original L15 parametrization 
is given by $2\ln K\approx 56$, which indicates a strong need for 
having $\zeta$ statistically. The median value of $\zeta\approx 0.064$ thus obtained 
implies that the characteristic redshift $z_c$ is $z\approx3.8$ for halos of 
$M_h(0)=10^{15}{\rm M_\odot}$, in comparison to $z_c\approx2.1$ for halos 
with $M_h(0)=10^{12}{\rm M_\odot}$. This increase of $z_c$ with host halo 
mass leads to flatter faint-end slopes for massive halos, giving  
better matches to the faint-ends of the CSMFs for both low-mass and high-mass 
halos. 

It is worth noting, however, that the uncertainties in the stellar mass estimates
may change the CSMFs in both the lowest and highest mass ends, where
the slopes of the CSMF are steep. As a test, we use only the CSMFs in the range
$M_*= [10^{8}, 10^{11}]{\rm M_\odot}$ as the observational constraints.  
In this case, the models with or without $\zeta$ are almost equally favored 
in terms of the Bayes factor. Given these, we conclude that
the original form of the L15 model can still accommodate the new CSMFs, 
and that the current data are still too uncertain to determine 
if a more extended model family is required.

\subsubsection{Model implementations}

We implement the empirical models described above to 
the dark matter halo population. We use the algorithm developed by 
\citet{parkinson08} to generate halo merger trees and to follow 
the build-up of dark matter halos. As mentioned above, this algorithm 
is based on the extended Press-Schechter formalism calibrated with results 
from $N$-body simulations. As shown in \citet{jiang14}, the predictions 
of this algorithm match accurately many properties of halo merger trees 
obtained directly from simulations, including 
halo mass assembly history, halo merger rate, and sub-halo mass functions.  

The empirical models described above also take into account some 
uncertainties in the observational data and in the model assumptions,
such as the intrinsic scatter in the stellar mass - halo mass relation, 
uncertainties in the stellar population synthesis and dust models, 
Eddington bias, and errors in redshift measurements. Unfortunately, how 
these uncertainties change as a function of redshift is poorly established. They are treated differently in different models. 
M13 adopted constant scatter in the stellar mass - halo mass relation and in the stellar mass estimate, 
while B13 parametrized the uncertainties as functions of redshift and 
treated them as a new set of free parameters to be determined in their 
model fitting. The treatment by Y12 lies in between. In our implementations,  
we follow each individual model as close as possible.  

We use WMAP7 cosmology to obtain the halo mass function, to 
construct halo merger trees, and to estimate distances from redshifts.
We adopt the \citet{chabrier03} IMF, the stellar population synthesis model 
of \citet{bruzual03} to account for stellar mass loss and to 
obtain stellar mass function from observations. These assumptions
are the same as adopted in the original models, except for Y12 where
a \citet{kroupa01} IMF was adopted. We correct the stellar masses of  
Y12 model by a factor of $\sim 1.4/1.7$ to match the IMF we adopt. 

\subsection{Hydrodynamical simulations}

We also test the predictions from two recent high-resolution, cosmological 
hydrodynamical simulations. The first is Illustris simulation 
\citep[][]{nelson15}, which follows $1820^3$ particles for each 
of the gas and dark matter components in a total volume of $(106.5\,{\rm Mpc})^3$, 
assuming  WMAP9 cosmology ($\{\Omega_m, \Omega_\Lambda, h\}=\{0.273, 0.727, 
0.704\}$). The other components the simulation traces are stars, stellar
wind particles, and super-massive black holes. 
The simulation starts from $z=127$ and includes physical 
processes such as radiative cooling, star formation, and various feedback processes. 
The free parameters in their model were constrained by using 
the star formation efficiency obtained from separate simulations 
that are more accurate in resolving small-scale structures. 
In our analysis, we use Illustris-1, their flagship simulation that
has the highest mass resolution ($1.6\times10^6 \rmn{M_\odot}$ and 
$6.3\times 10^6\rmn{M_\odot}$ for baryon and dark matter, respectively). 
To match the set of observations adopted here for model testing,  
we use the snapshot at $z=0.03$, which contains a total of $7,647,219$ 
groups identified by the FoF algorithm. In the simulation, galaxies are
defined according to the spatial distribution of stars and stellar wind particles, 
and the brightness profile fit to them. The simulation assumes 
the \citet{chabrier03} IMF and the stellar population synthesis model 
of \citet{bruzual03}. As the cosmological parameters of WMAP9 are similar
to those of WMAP7, the difference in cosmology is ignored
in our analysis. 
We bin their stellar masses to obtain the stellar mass function (SMF).

Another simulation we use is the Evolution and Assembly of GaLaxies and 
their Environments \citep[EAGLE;][]{schaye15}. EAGLE
traces the evolution of gas, stars, dark matter, and 
massive black holes, and implements physically motivated models for 
gas cooling, star formation law, stellar and AGN feedback. The free 
parameters of the feedback models were tuned to match the 
SMF and black hole mass - stellar mass relation at $z\sim 0$. 
The simulation starts from $z=127$ and adopts cosmological parameters 
from {\it Planck}: $(\Omega_m, \Omega_\Lambda, h)=(0.307,0.693,0.678)$
\citep[][]{planck14}. We use their simulation of the largest volume of 
$(100{\rm Mpc})^3$ for our analysis. It contains $\sim 10,000$ galaxies 
with stellar masses similar to or above that of the Milky Way.
Unfortunately, recalibrating their result to account for different
cosmology is not trivial, since the impact of changing the parameters
to the mass function is highly non-linear in principle. However, 
the other uncertainties that enter the models or the data must 
overpower the change in cosmology. We thus do not attempt
any recalibration of the simulation results to account for the difference in cosmology. 
The \citet{chabrier03} IMF and the spectral synthesis model of 
\citet{bruzual03} were assumed. We bin their stellar masses to get 
the stellar mass function.

\section[Data]{Comparisons with observational data}
\label{sec_data}

In this section, we describe all observational data used for our analyses. 
These include the stellar mass function (SMF) of the general 
galaxy population (\S3.1), the SMFs separately for central and 
satellite galaxies (\S3.2), and the conditional stellar mass functions
(CSMFs) of galaxies (\S3.3).

\subsection{The field stellar mass function of galaxies at $z\sim 0$} 

\begin{figure*}
\includegraphics[width=0.9\linewidth]{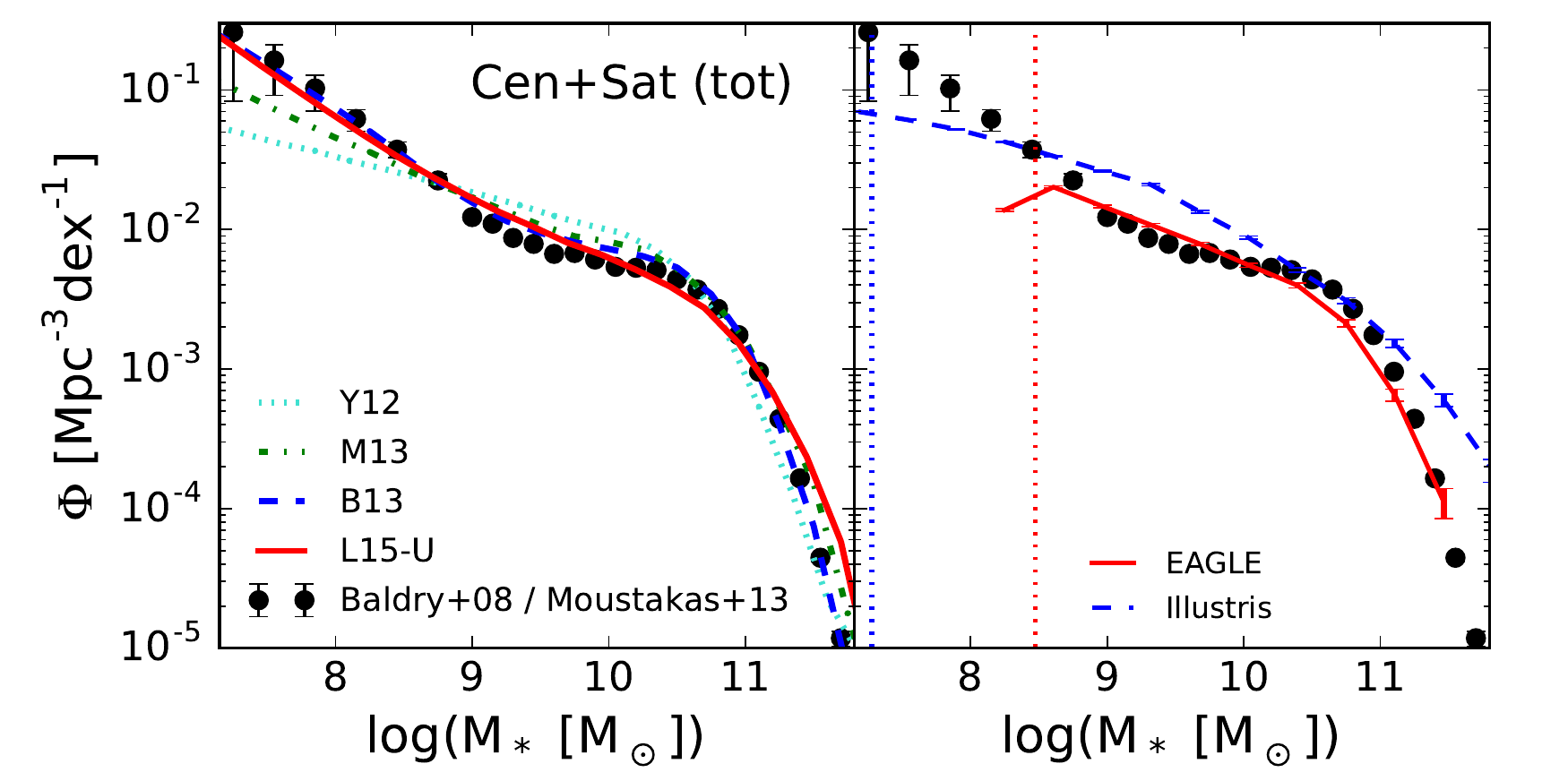}
\caption{The observed stellar mass function of galaxies (data points) in comparison
with the predictions of individual empirical models (left) and hydrodynamical 
simulations (right), 
as indicated in the panels. The vertical lines in the right panel show the 
resolution limits of the two simulations, as given in the original papers 
describing the simulations. The Poisson errors are presented for the simulations.}
\label{fig_LF_local_model}
\end{figure*}

We use the local ($z\approx0.1$) SMF obtained from the combination 
of the results obtained by \citet{baldry08} and \citet{moustakas13}.
The data for stellar masses below $M_*\approx 10^{9}{\rm M_\odot}$ is from 
Baldry et al., while the data at larger stellar masses is from Moustakas et al..
Here we briefly summarize the methodologies with which the SMFs were computed, and 
refer the reader to their original papers for details. 

Baldry et al. used the New York University Value-Added Galaxy 
Catalogue \citep[NYU-VAGC;][]{nyu}, 
which includes $49,968$ galaxies at $z<0.05$, to construct the local SMF.
They adopted the stellar mass estimates from \citet{kauffmann03}, 
\citet{gallazzi05}, and \citet{panter07}. In the data set we use, 
the stellar mass estimates are corrected to a \citet{chabrier03} IMF. 

Moustakas et al. estimated the local SMF using galaxies cross-identified between 
the Sloan Digital Sky Survey Data Release 7 \citep[SDSS DR7;][]{dr7} and the 
{Galaxy Evolution Explorer} \citep[GALEX;][]{galex} Deep
Imaging Survey. This results in $\sim 170,000$ galaxies with a total sky coverage
of $2505\ {\rm deg}^2$. Near-infrared photometry of these galaxies was obtained
from the Two Micron All Sky Survey \citep[2MASS;][]{2mass} and the 
{Wide-field Infrared Survey Explorer} 
\citep[WISE;][]{wise}. 
The photometry in a total of 12 bands (near-UV and far-UV of the 
{GALEX}, {\it ugriz} bands from SDSS model magnitudes, $JHK_s$ magnitudes from the 2MASS, 
and the integrated photometry at $3.4$ and $4.6\micron$ from the 
{WISE} All-Sky Data Release) 
was used to infer the galaxy stellar masses from spectral energy 
distribution (SED) modelling. More specifically, 
Moustakas et al. used the Flexible Stellar Population Synthesis model of 
\citet{conroy09}, a \citet{chabrier03} initial mass function (IMF), 
exponentially declining star formation histories (SFHs), and the dust 
attenuation curve of \citet{charlot00}, to model the SEDs of individual galaxies.
The SMF obtained by \citet{moustakas13} is in good 
agreement with some previous measurements, such as those of \citet{cole01},
\citet{li09}, and \citet{baldry12}. See their Appendix B 
for detailed analyses how variations in the IMF, SFH, spectral synthesis model, 
and dust attenuation can affect the SMF obtained.  

The left panel of Figure~\ref{fig_LF_local_model}
compares the predictions of the empirical models with the observed local SMF 
described above. As one can see, the prediction of the Y12 model 
is too flat in the low-mass end to match the upturn 
seen in the observation. This discrepancy owes partly to 
the simple functional form (a double power-law) they adopted 
for the stellar mass-halo mass relation, and partly 
to the SMFs that they used as observational constraints.
In fact, Y12 found that the two sets of 
SMFs at high redshifts that they adopted led to significant 
differences in the inferred values of model parameters. The results used 
here are the predictions of `SMF2' referred in the original paper. 
The model of M13 also predicts a shallower faint-end slope than the observational
data. Similar to Y12, M13 also adopted a simple double power-law form for the 
stellar mass - halo mass relation, and used a local SMF that has shallower 
faint-end slope than the one adopted here to constrain their parameters. 
In contrast, the prediction of B13 matches well the observed SMF, even 
in the faint end. B13 adopted a rather flexible functional form 
for the stellar mass - halo mass relation, which is probably required 
to match the faint-end upturn in the SMF. In addition B13 adopted the combined SMF 
of \citet{baldry08} and \citet{moustakas13} as one of their 
observational constraints, and so the good match between the model prediction and the data is not 
surprising. The prediction of the L15 model also matches well the 
observational data. Note that L15 used the SMF of \citet{baldry12} as an 
observational constraint. Their SMF extends only to $10^{8.5}{\rm M}_\odot$
and so the faint-end upturn is not well represented. The faint-end upturn 
predicted by L15 is largely due to the CSMF 
of galaxies in rich clusters, as given by \citet{popesso06}, they 
adopted to constrain their model. 

The right panel of Figure~\ref{fig_LF_local_model}
compares the numerical simulation results with the 
observational data. Illustris simulation produces too many 
galaxies in the intermediate mass range as well as in the massive end, 
but too few low-mass galaxies. The overall shape of the predicted SMF 
is very different from that of the observed SMF. On the other hand, 
the prediction of EAGLE simulation matches the observational data 
reasonably well above the resolution limit.  This may not be very surprising, 
because the free parameters in EAGLE simulation were tuned to match 
local observations. Unfortunately, the relatively poor mass resolution does 
not allow us to investigate whether a faint-end upturn is predicted 
in the simulation.

\subsection{Central and satellite galaxies} 

\begin{figure*}
\includegraphics[width=0.9\linewidth]{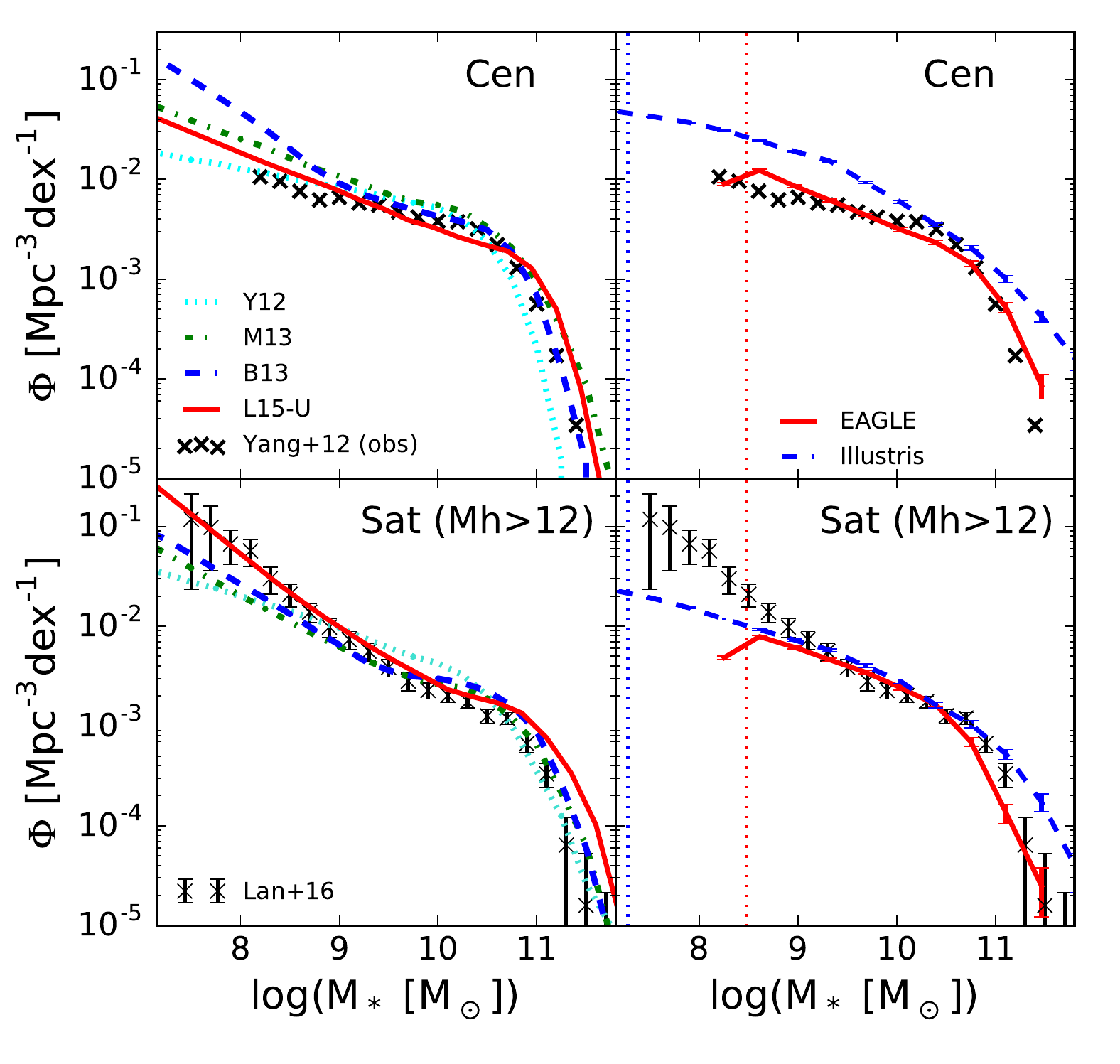}
\caption{The observed stellar mass function of central (data points in the upper 
two panels) and satellite (data points in the lower two panels) galaxies,
in comparison with the predictions by individual empirical models (left panels) 
and gas simulations (right panels), 
as indicated. The completeness in stellar mass from the observation of centrals 
is not guaranteed for $M_*<10^8 {\rm M\odot}$. The vertical lines in the right 
panels show the resolution limits of the two simulations, as given in the 
original papers describing the simulations.}
\label{fig_LF_local_cs}
\end{figure*}

 Using the group memberships provided by \citet{yang07}
group catalog (see next subsection for more details), we can 
separate galaxies into two populations, centrals and satellites. A central galaxy is
defined to be the most massive member in a group, while all other members 
in a group are called satellites. The CSMFs can then 
be estimated separately for the centrals and satellites. Formerly the total 
SMF can be expressed in terms of these conditional functions as  
\begin{eqnarray}\label{eq_smf}
\Phi_{\rm tot}(M_*) &=& \int_{M_{h,min}}^{\infty}dM_h\ n(M_h)\times \\\nonumber
&&\qquad \qquad \left\{\Phi_{\rm cen}(M_*|M_h)+\Phi_{\rm sat}(M_*|M_h)\right\}, 
\end{eqnarray}
where $\Phi_{\rm cen}(M_*|M_h)$ and $\Phi_{\rm sat}(M_*|M_h)$ are the CSMFs 
of the centrals and satellites, respectively, in halos of mass $M_h$, 
while $n(M_h)$ is the halo mass function, which is the number density of 
halos of masses between $M_h$ and
$M_h+dM_h$. In \citet{lan16}, the CSMFs 
are given only for satellites in groups with halo masses above $10^{12}{\rm M}_\odot$
(see the next subsection for details). The satellite 
SMF used here is obtained directly from their measurements by summing up 
the CSMFs of such halos. For central galaxies,  we use 
the results obtained by \citet{yang12} from their group catalog. Since the group catalog is 
based on the SDSS spectroscopic data, the central SMF was measured only for galaxies 
above $10^{8}{\rm M}_\odot$ (see table 6 in their paper).

The data points in Figure~\ref{fig_LF_local_cs} show the SMFs for central and satellite 
galaxies, respectively. Separating galaxies into centrals and satellites provides 
more information about the galaxy population than the total SMF alone, and 
Figure~\ref{fig_LF_local_model} and Figure~\ref{fig_LF_local_cs} 
demonstrate this point clearly. For instance, although the empirical 
model by B13 (see \S\ref{sec_model}) matches well the faint-end upturn 
in the observed total SMF, this match is now revealed as due to an excess in 
the SMF of central galaxies combined with a deficit in the SMF of satellite galaxies.
The M13 model has similar problems; it under-estimates the number of satellite galaxies
at the low-mass end even more strongly than B13. The Y12 model matches the central SMF 
reasonably well, but it fails to reproduce the strong upturn in the low-mass end seen 
in the observed SMF of satellite galaxies. Overall, the L15 model can match both the 
observed central and satellite SMFs, although some discrepancies in details can still be 
seen. This match is not trivial, because these observations were not used 
as constraints in L15.

The comparisons of the two gas simulations with the observational results are shown 
in the right two panels of Figure~\ref{fig_LF_local_cs}. Here we see that the 
EAGLE simulation matches the observational data reasonably well above its mass 
resolution limit. Illustris simulation matches the SMF of satellites only in 
the intermediate mass range; it over-predicts the central SMF over almost the 
entire mass range, except at the knee of the SMF.  

\subsection{The conditional stellar mass functions of galaxies in groups} 

\begin{figure*}
\includegraphics[width=0.9\linewidth]{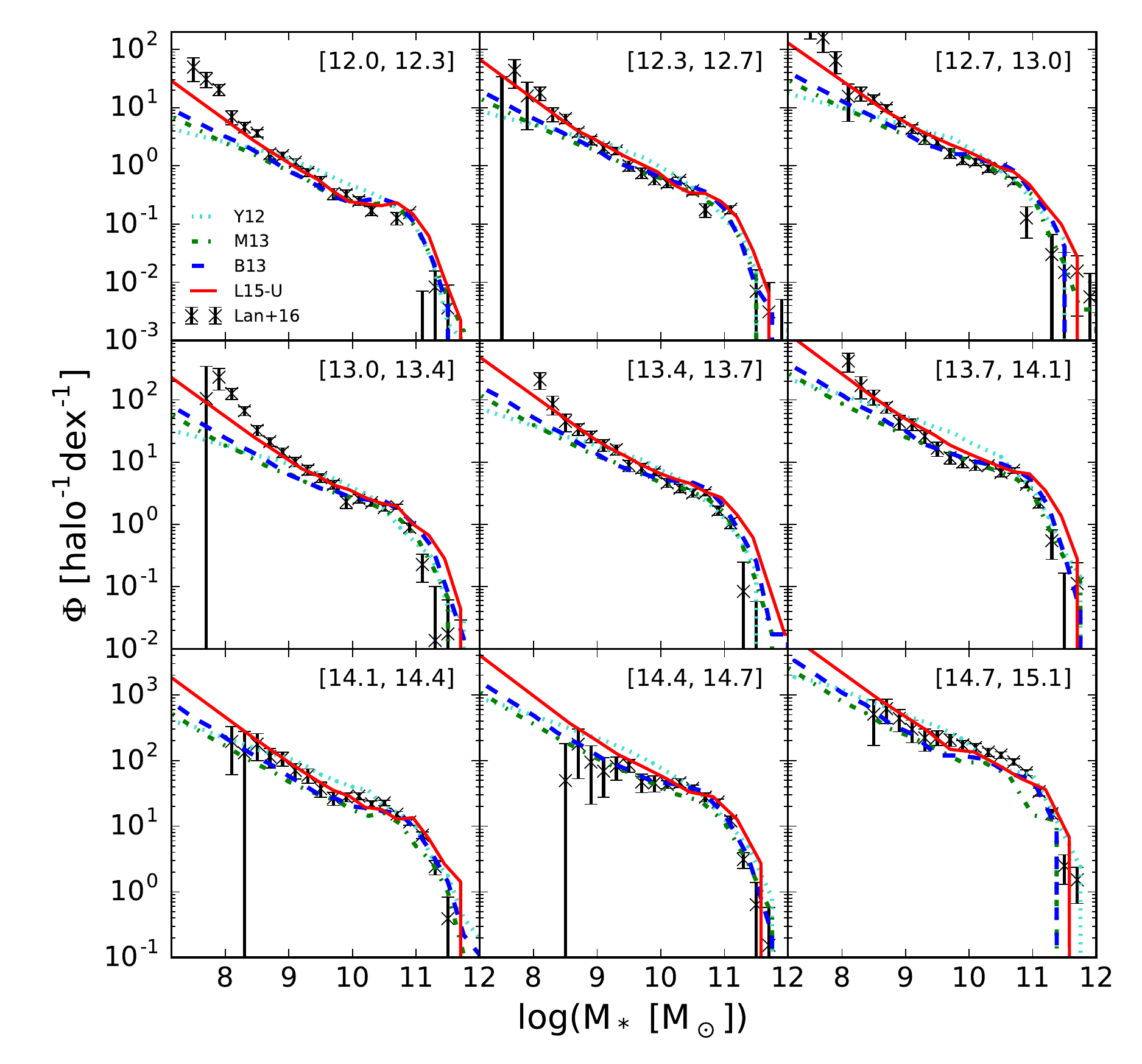}
\caption{The observed (data points) and predicted (lines) conditional 
stellar mass functions of galaxies in groups of different halo masses, 
as indicated in individual panels.}
\label{fig_CLF_model}
\end{figure*}

\begin{figure*}
\includegraphics[width=0.9\linewidth]{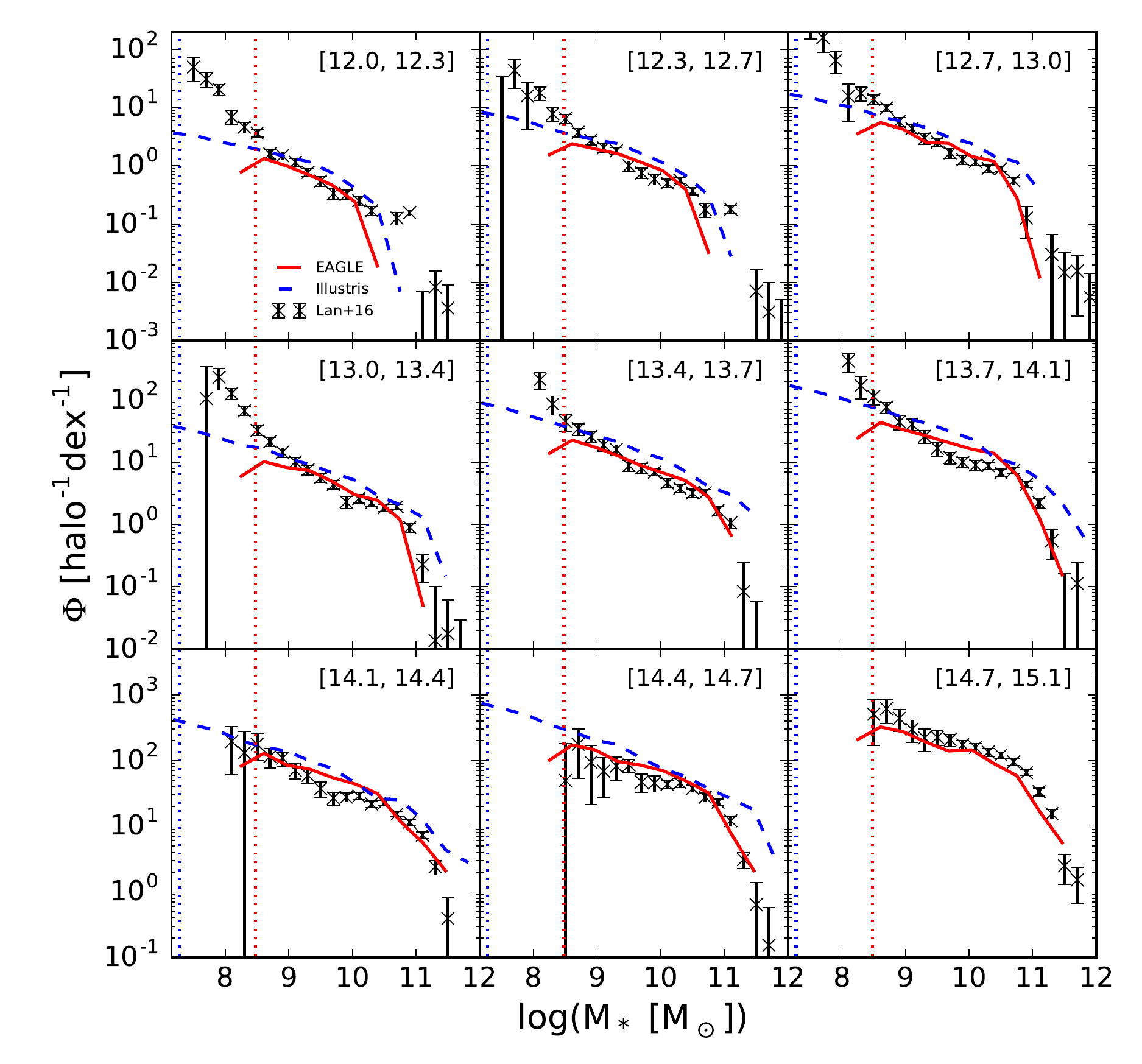}
\caption{The comparison between the observed conditional stellar mass 
functions of galaxies (data points) with the results of EAGLE and Illustris
simulations (lines), for groups of different halos masses, 
as indicated in each panel. The two vertical lines indicate the mass limits 
of the two simulations.}
\label{fig_CLF_simul}
\end{figure*}

We use the CSMFs
obtained by \citet{lan16} as our main 
data set to compare with models. Here we summarize briefly their methodology and results. 
Lan et al. used galaxy samples from the NYU-VAGC, 
which is based on the Sloan Digital Sky Survey 
Data Release 7 \citep[SDSS DR7;][]{dr7}. A $K$-correction was applied 
using the model of \citet{blanton03}. In order to associate galaxies with 
clusters/groups of galaxies, they adopted the group catalog of \citet{yang07}, 
which was constructed by applying the halo-based group finder developed
by \citet{yang05} to the SDSS DR7. The group finder assigns galaxies 
into halos using certain criteria in phase space, and galaxies residing in 
a common halo are considered to be members of the same group. More specifically, 
a tentative halo mass is assigned to a tentative group based on
the galaxies that have already been assigned to the group, assuming 
a monotonic relation between the total stellar mass of all 
assigned members with $M_r<-19.5$ and halo mass. The tentative mass is then 
used to estimate the virial radius and velocity dispersion of the 
halo, which in turn are used to update the group membership. The 
procedure is iterated until both group memberships and halo masses 
converge for all groups. \citet{yang07} used mock catalogs constructed 
from $N$-body simulations to show that the dark matter halo masses 
estimated in this way are consistent with those directly obtained from the 
simulations, with scatter of $\sim 0.3$ dex over three orders
of magnitude in halo masses that cover the mass range relevant to our analyses. 
When we compare the CSMFs obtained from models with the observational 
results, an uncertainty of $\sim 0.3$ dex is included in the model 
predictions. The halo masses used here  are $M_{200}$, the total mass 
enclosed by a radius, $r_{200}$,  within which the average density is $200$ times 
the mean density of the universe. 

Lan et al. used only groups at $z<0.05$, 
where halos with masses of $M_{200}>10^{12}{\rm M_\odot}$ are complete. 
To limit the uncertainty in redshifts due to peculiar velocities, they 
also eliminated groups at $z<0.01$. With the groups and their positions 
identified in the SDSS DR7 survey area, Lan et al. estimated the 
excess of galaxy number in each luminosity bin within a projected 
distance of $r_{200}$ of each group. The conditional luminosity function (CLF) of galaxies is then 
obtained by averaging galaxy counts within all groups of a given halo 
mass, with subtractions of the background and projection effects due 
to clustering on large scales \citep[see][for the detail]{lan16}. 
Lan et al. applied this method to the photometric sample of SDSS DR7, 
down to a $r$-band model magnitude of $21$. This corresponds to $M_r\approx-12$ 
(or $L\approx10^7L_{\odot}$) and $M_r\approx-14$ at $z=0.01$ and $0.05$, respectively. 
Since the number density of more massive halos is smaller, Lan et al. 
was able to estimate the CLF down to $M_r\sim-12$ for 
low-mass halos ($M_{200}\sim 10^{12}{\rm M_\odot}$), but only to  
$M_r\sim-14$ for massive halos. 

To convert their CLFs into the corresponding CSMFs, 
we use a mass-to-light relation based on galaxy colors and luminosities \citep[e.g.][]{bell03} 
to obtain the stellar masses of galaxies. However, the uncertainty in 
the observed galaxy colors, especially for faint galaxies, may bias the stellar mass 
estimates and, therefore, the stellar mass functions. To reduce such bias, we first separate 
galaxies into blue and red populations by using the $u$-$r$ color separation 
suggested by \citet{baldry04} [see their equation (11)]. We then use the observed 
luminosity of a galaxy and the mean $u$-$r$ color for the galaxy population at that 
luminosity, instead of the observed color of the galaxy, to 
estimate the stellar mass. The mean $u$-$r$ color-luminosity relations for the blue and red 
populations are derived in \citet{lan16} [their equations (C2) and (C3)] based 
on the same data set. We have made tests either by using the observed galaxy color or
by artificially introducing some uncertainties in the galaxy color, and found  
that all these do not lead to any qualitative change of our results. 
Note again, as described in \S\ref{sec_model}, our model predictions for 
the stellar masses of individual galaxies also include some uncertainties 
in the stellar mass estimates to mimic the uncertainties in the observational 
stellar masses. Lan et al. adopted a \citet{kroupa01} IMF for the CSMFs.

With the estimated stellar masses of individual galaxies, we measure the 
CSMFs using the same method Lan et al. did for the CLFs. The stellar mass functions 
are measured down to the limiting stellar masses at which both the stellar masses of 
blue and red galaxies derived from the flux limit photometric sample ($r<21$) are complete. 
In addition, the limiting stellar mass bins are selected to ensure that 
they each contain at least five groups. We bootstrap the group catalog 200 times to 
estimate the errors in the derived CSMFs. 

\begin{figure*}
\includegraphics[width=0.95\linewidth]{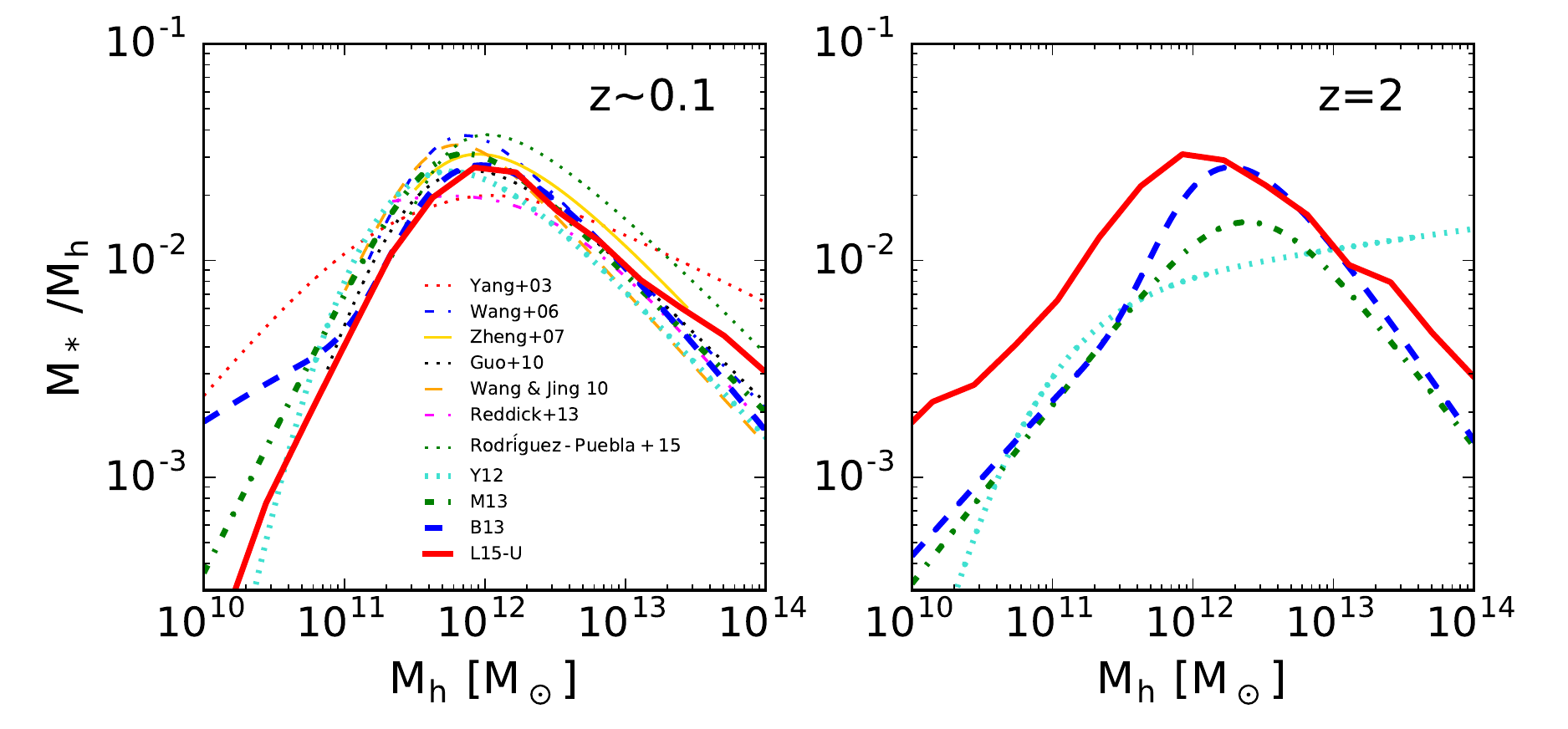}
\caption{The average stellar mass - halo mass relation for 
central galaxies from the empirical models considered 
in this paper, along with the results in the literature 
from recent studies that adopted empirical approaches such as 
halo abundance matching, conditional luminosity function 
and halo occupation distribution, for local Universe (left) and $z=2$ (right).}
\label{fig_MsMh}
\end{figure*}

Figure~\ref{fig_CLF_model} compares the CSMFs to 
those predicted by the empirical models. The predictions by the M13 and B13 
models are qualitatively similar, with B13 predicting more low-mass galaxies.
Both models under-predict the CSMFs at the low stellar mass ends, and 
the under-prediction is more significant for groups of lower halo masses. 
Only for massive clusters are the predictions consistent with the observational data. 
The predictions of Y12 are too shallow in the low mass end; the model systematically 
under-predicts the CSMF at the low mass end and over-predicts that in the 
intermediate mass range. In particular, Y12 does not predict any upturn seen in the data.

The L15 model matches the overall behaviors of the CSMFs over the entire halo mass range. 
It also matches the CSMFs in detail for most of the halo mass bins.
However, the low-mass upturn it predicts for more massive halos may be 
too steep, especially for the two most massive samples. As mentioned above, 
the L15 model used the composite CLF of galaxies in rich clusters
given by \citet{popesso06} as one of the constraints on their model. 
The faint-end upturn in this composite CLF is significantly steeper than 
that of Lan et al. used here. The over-prediction is therefore due to the 
observational data which the model was tuned to match with. 
The model seems to under-predict the CSMF at the low-mass 
end in two mass bins: the lowest mass bin of 
$\log(M_h/{\rm M}_\odot)=[12.01, 12.34]$, 
and the intermediate mass bin of 
$\log(M_h/{\rm M}_\odot)=[13.03, 13.37]$. It is unclear if these
discrepancies are due to random fluctuations in the data, 
or indicate that the L15 model has to be modified to accommodate 
the data. We will come back to this in the following section.

Figure~\ref{fig_CLF_simul} compares the observational data with the two 
gas simulations. Illustris simulation mismatches the observation
over a wide range of stellar masses for almost all the halo mass bins. Overall, 
the simulation significantly under-predicts the CSMFs at the 
faint ends, and over-predicts them in both the massive and intermediate mass 
ranges. EAGLE simulation appears to be in better agreement with the 
observation, except that it does not reproduce sufficient number of massive galaxies 
in low-mass halos. Unfortunately, its mass resolution prevents us 
from probing its behavior at the faint end.


\section{Model predictions}
\label{sec_impl}

In this section, we compare all the empirical models in their 
predictions for the stellar mass - halo mass relation, the star 
formation rates and stellar masses in halos of different masses at 
different redshifts, and for the SMFs of high-redshift galaxies. 

\begin{figure*}
\includegraphics[width=0.95\linewidth]{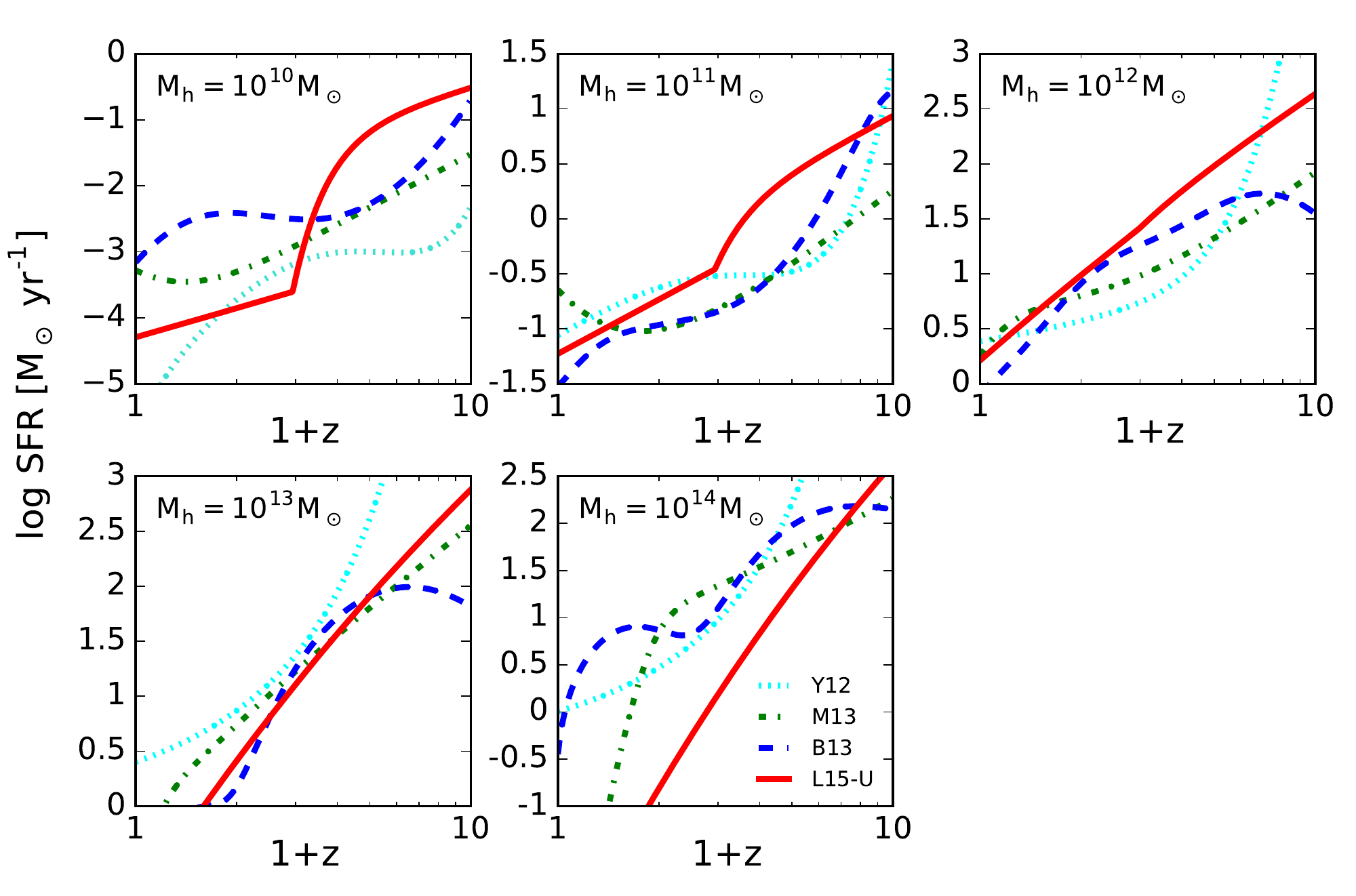}
\caption{The average star formation rate of central galaxies as a function 
of redshift for halos of different masses, as predicted by 
various empirical models, as indicated.}
\label{fig_SFR_model}
\end{figure*}

\begin{figure*}
\includegraphics[width=0.95\linewidth]{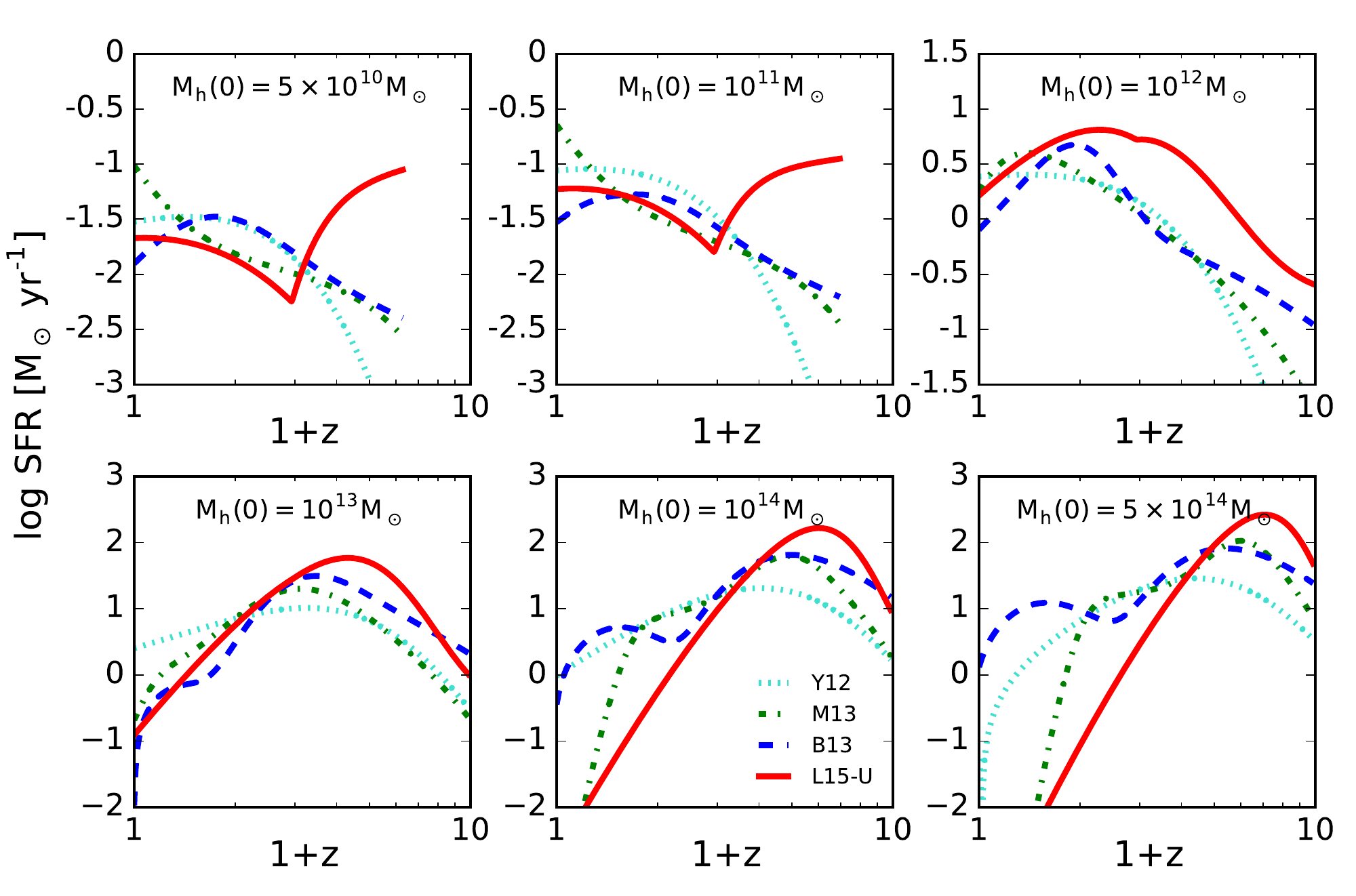}
\caption{The average star formation rate of central galaxies as a function 
of redshift for halos of different present-day masses, predicted by 
the empirical models, as indicated. }
\label{fig_SFR_Mh0}
\end{figure*}

\subsection{Stellar mass - halo mass relation}

Figure~\ref{fig_MsMh} shows the stellar mass - halo mass relation predicted 
by our updated model, in comparison with the predictions of the 
other three empirical models considered here and the results from 
the literature. Different models made different assumptions for 
conversions from luminosity to stellar mass, for prescription 
of scatter in the relation, and cosmological models. They also
employed different observations as constraints for the models. 
Given all these differences, it is remarkable that the predictions 
of most models are consistent with each other within $\sim0.2$ dex 
at $z\sim 0.1$ for a large range of halo masses.  
All models predict a characteristic mass scale, $M_h\sim 10^{12}{\rm M\odot}$, 
at which the stellar mass to halo mass ratio peaks. 
Among the more recent results, B13 is an exception in that it predicts 
a strong upturn at the low-mass end. The earlier result of \citet{yang03} 
was obtained by using their luminosity - halo mass relation together 
with the assumption of a constant stellar mass to luminosity ratio, 
$M/L=1.8\ {\rm M_\odot}/L_\odot$ (in the $b_J$ band of 2dFGRS which 
they used to constrain their model). 

At higher redshift, however, the predictions by different empirical models 
differ significantly. In particular, the update of L15, L15-U, predicts
a much higher star formation efficiency for low mass halos at high redshift, 
because of the boost of star formation rate at $z>z_c$ in low mass halos 
to match the upturns in the CSMFs.

\begin{figure*}
\includegraphics[width=0.95\linewidth]{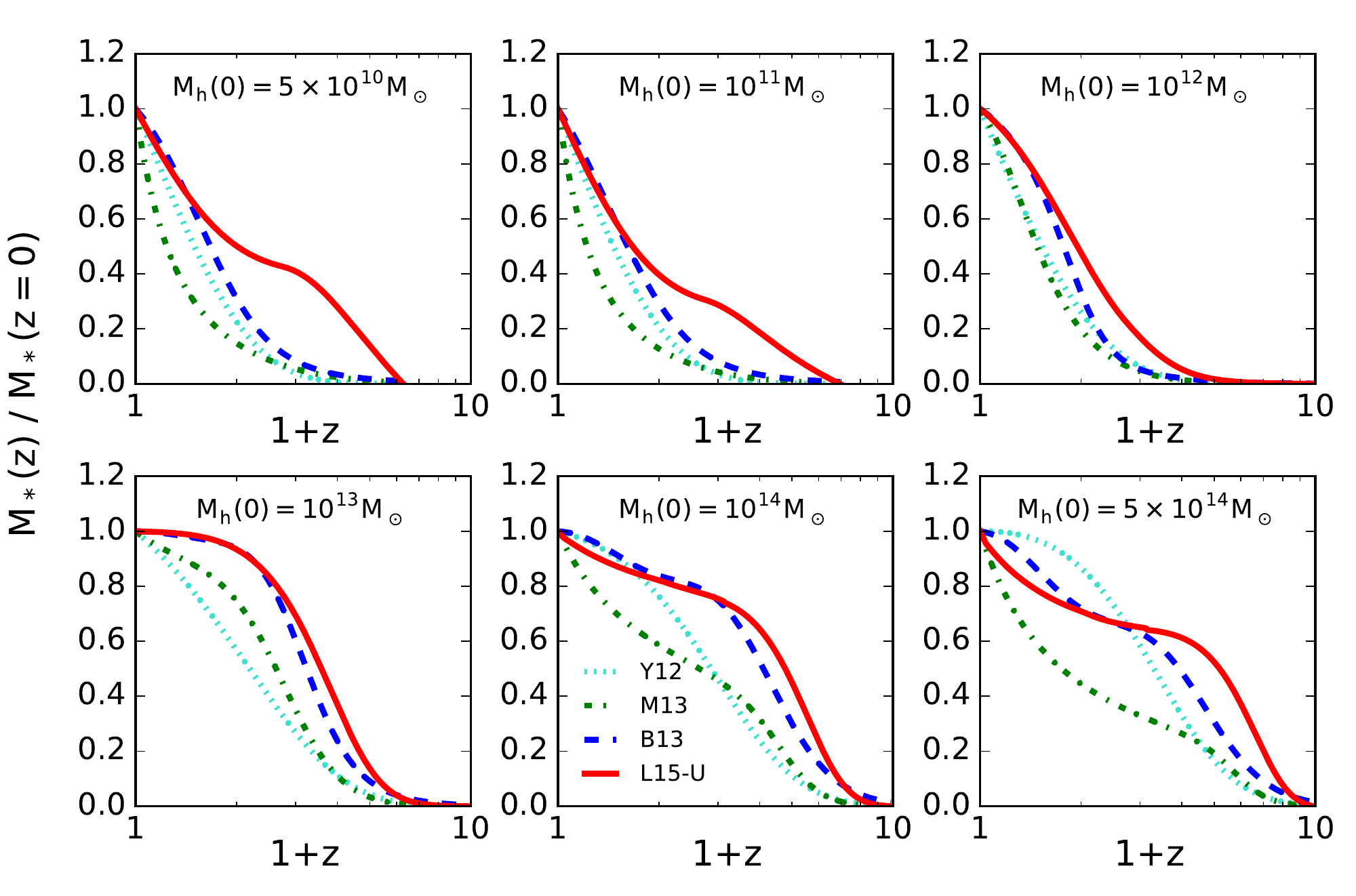}
\caption{The average stellar mass assembly history of central galaxies 
as a function of redshift for halos of different present-day masses, 
predicted by various empirical models as indicated. }
\label{fig_assembly}
\end{figure*}

\subsection{Star formation histories in dark matter halos} 

Figure~\ref{fig_SFR_model} compares the empirical models in terms of their
predictions for the average star formation rate (SFR) of central galaxies in halos at 
different redshifts. Some models predict complicated star formation histories
that are clearly due to over-fitting of the observational data. 
The L15-U model predicts much higher SFRs at $z\ge 2$ in low-mass halos 
than other models, which is clearly a consequence of the strong upturns at the 
faint-ends of the CSMFs used to constrain the model. 
The existence of a characteristic redshift, $z\sim 2$, is clearly seen   
in halos with $M_h<10^{12}\,{\rm M}_\odot$, and its physical implications 
will be discussed later. 

Figure~\ref{fig_SFR_Mh0} shows the model predictions for the average star formation 
histories of central galaxies in halos of different present-day masses. 
The predictions of different models are very different. In particular, 
for present-day dwarf galaxies that reside in halos of $M_h(0)<10^{11}{\rm M_\odot}$, 
L15-U predicts a very active star formation episode at $z>2$. In contrast, 
most of the stars in such halos are formed at $z<2$ in all other models. 
This difference has other observational consequences. Indeed, as discussed 
in \citet{lu14} and \citet{lu15}, the early starburst in low-mass halos 
predicted by L15 is consistent with the observations that a significant 
fraction of old stellar population exists in local dwarf galaxies  
\citep[e.g.][]{weisz11} and that the star formation rate function 
at the low-rate end is very steep at $z>4$ \citep[e.g.][]{smit12}.
For Milky-Way sized halos, the star formation history predicted 
by L15-U is broader than those predicted by the other three models. 
For massive halos with $M_h(0)\ge 10^{14}{\rm M_\odot}$,   
L15-U predicts a decline of the SFR with decreasing 
redshift starting from relatively high redshifts, in contrast 
to the predictions of B13 and Y12 that the star formation rates 
remain relatively high all the way to the present time, and to 
the prediction of M13 that a rapid decline only occurs at 
$z<1$. 

All these results demonstrate that different empirical models 
can make vastly different predictions for the star formation histories 
for present-day galaxies, even though all the models 
are tuned to match the observed SMFs.

\subsection{Stellar mass assembly histories}

Figure~\ref{fig_assembly} shows the average stellar mass assembly histories 
for the central galaxies in halos of different present-day masses predicted by 
different models. The model predictions take into account {\it in situ} star 
formation, accretion of satellites, and stellar mass loss due to stellar 
evolution. Again, for low-mass halos, where the increase of stellar mass 
is dominated by {\it in situ} star formation \citep[][]{lu15}, L15-U is 
distinct from the other models in that about half of their  
stellar mass at the present was already in place by $z\sim2$ 
via star formation (see Figure~\ref{fig_SFR_Mh0}). 

For Milky-Way sized halos, however, the differences between the model 
predictions are milder. All the models predict that  
about half of stellar mass was in place by $z\sim1$. There is a significant 
difference between L15-U and other models at high $z$. For example, 
L15-U predicts that about 15\% of the final stellar mass was assembled by 
$z\sim 2$, while less than 10\% was predicted by the other models.   

For central galaxies in present-day massive halos with 
$M_h(0)>10^{14}{\rm M}_\odot$, the predictions of different models 
again become very different. M13 predicts a much later 
assembly for these galaxies than any other models. The predictions 
of B13 look similar to L15-U, but the increase in stellar mass
with time is due to different reasons. While L15-U predicts that 
the increase at $z<2$ is dominated by accretion of stars from satellites, 
B13 predicts that a significant fraction of the increase at $z<2$ 
is actually due to {\it in situ} star formation (see Figure~\ref{fig_SFR_Mh0}). 
This difference is again due to the boost of star formation 
in low-mass halos at high $z$ in the L15-U model. The increased 
amount of stars formed in progenitors at high $z$ makes the accretion 
of stars more important in the growth of stellar mass in 
a massive galaxy, and the fraction of stars formed in situ 
has to be decreased proportionally in order to match the final 
stellar mass of the galaxy. The results demonstrate the importance 
of properly modeling the star formation in low-mass progenitors 
at high $z$ in order to understand the star formation and stellar mass 
assembly histories of massive galaxies at the present day.

\subsection{Stellar mass functions of high-redshift galaxies} 

\begin{figure*}
\includegraphics[width=0.9\linewidth]{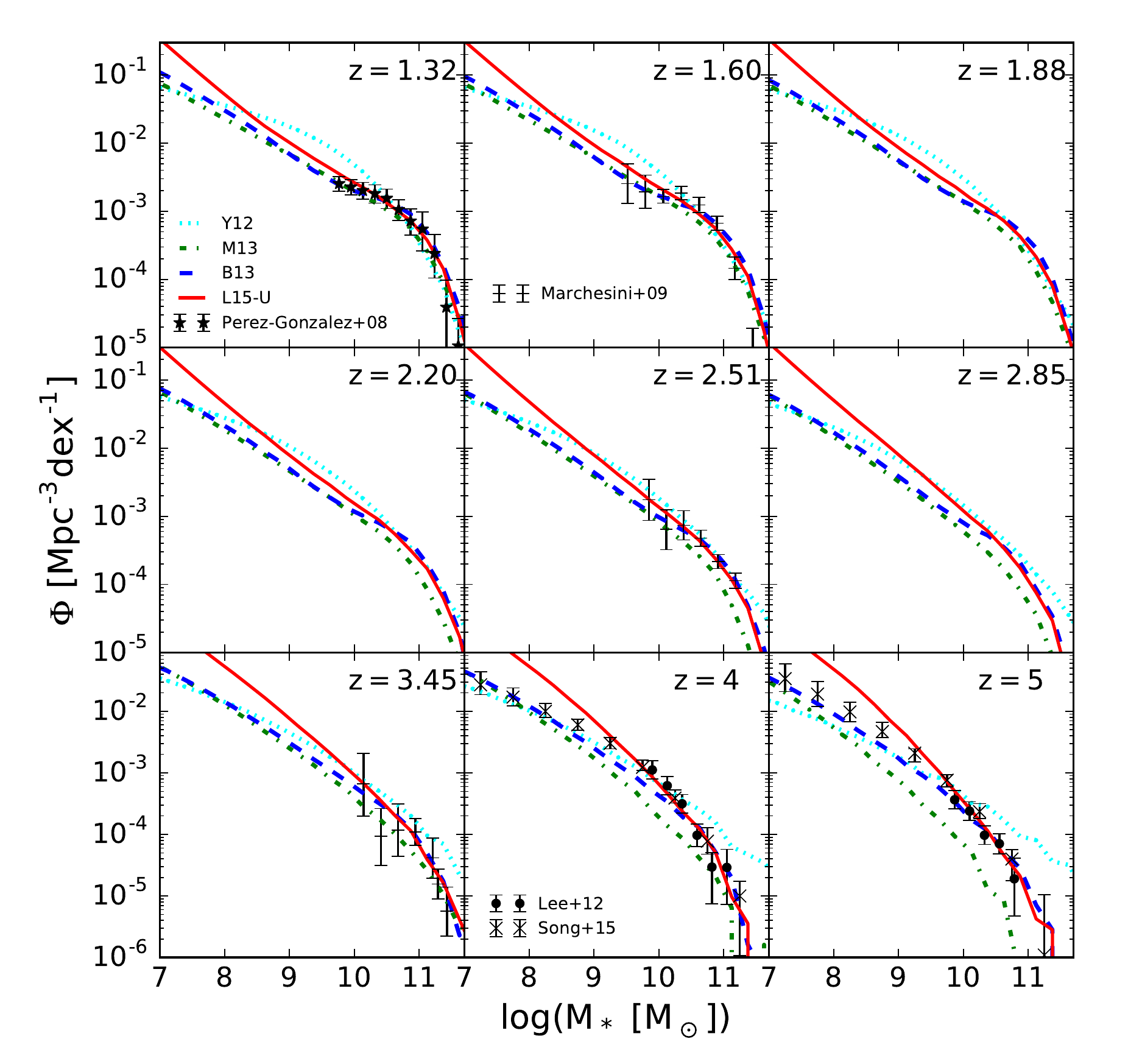}
\caption{The model predictions for the field stellar mass functions at high 
redshifts (solid) in comparison with observations. }
\label{fig_highz} 
\end{figure*}
 
Figure~\ref{fig_highz} shows the predictions of the empirical models 
for the stellar mass functions of galaxies at a number of redshifts. 
The predictions of B13 and M13 are similar in both slopes and 
amplitudes at the low mass ends, but B13 predicts many more 
massive galaxies than M13, particularly at high redshifts.
Y12 predicts significantly flatter slopes at the low-mass ends, 
and more galaxies in the intermediate mass range, than the other three 
models. The stellar mass functions predicted by the L15-U model 
match the predictions of B13 at $M_*>10^{10}{\rm M}_\odot$, 
but are significantly steeper at the low-mass ends.

We select some observational SMFs of high redshift galaxies 
from the literature to compare with the model predictions. Specifically, we use the SMFs at 
$1.3<z<3.5$ given by \citet{pg08} and \citet{marchesini09}. 
P\'erez-Gonz\'alez et al. used a sample combining data in three different fields 
with a total area of $664\ {\rm arcmin}^2$ that have a total $\sim 28,000$ systems 
selected with the $3.6-4.5\micron$ photometry of {\it Spitzer Space Telescope} 
\citep[][]{werner04}. The sample is complete down to $M_*=10^{10}{\rm M_\odot}$.
Marchesini et al. combined data from the deep NIR MUSYC, the ultra-deep 
FIRES, and the GOODS-CDFS surveys to derive the SMFs from the optical to MIR 
broad bands photometry. P\'erez-Gonz\'alez et al. assumed a \citet{salpeter55} IMF while 
Marchesini et al. adopted a pseudo-\citet{kroupa01} IMF. It is known that the 
stellar mass estimated using a Salpeter IMF is roughly a factor of 
1.4 higher than that given by a pseudo-Kroupa or Chabrier IMF, and we correct 
all the stellar masses to the IMF we adopt here. For the SMFs at even higher 
redshifts, $z=4-5$, we use the results by \citet{lee12} and \citet{song16}.
Both B13 and L15-U match the observational data well at 
$M_*>10^9 {\rm M}_\odot$, while the other two models match the high-$z$ data 
poorly at the high-mass end. The prediction of the L15-U is significantly 
steeper than the observational results given by Song et al. at $z>4$. 
If the high-$z$ SMFs are as shallow as those given by Song et al., then 
there may be a tension between the observed CSMFs at low $z$ and the 
observed SMFs at high $z$, at least within the model family represented 
by the halo-based empirical model of L15. 

\section{Summary and discussion}
\label{sec_sum}

Galaxy formation and evolution within the current cosmological frame 
are controlled by a number of physical processes, many of which are 
still poorly understood from first principles. In the absence of a
proper understanding of these processes, halo-based empirical models 
provide a useful way to establish the link between galaxies and CDM halos 
purely on the basis of observations and the current cosmology.   
In this paper we use a variety of galaxy stellar mass functions to test 
a number of popular empirical models. In particular, 
we focus on using the conditional stellar mass functions (CSMFs) of galaxies 
in galaxy groups as obtained by \citet{lan16} to test the models. 
We find that the CSMFs predicted by different models can be very different, 
even though they are all tuned to match the observed stellar mass function 
of the total galaxy population. This clearly demonstrates the power of 
the CSMFs in constraining models. Since the CSMFs are 
measured from observations in the nearby Universe, the samples 
that can be used are larger, and the stellar mass functions can be 
measured to the low-mass ends. As the galaxies that reside in present-day 
galaxy systems, such as clusters and groups of galaxies, are expected 
to have formed at various redshifts, the CSMFs in groups/halos of 
different masses carry important information about galaxy formation 
in dark matter halos at different redshifts.  

The CSMFs are then used as constraints to update the original model by \citet{lu14,lu15}. 
The model parameters obtained here are very similar to those obtained in the original 
paper which uses a completely different set of observational constraints, demonstrating 
that the different data sets are consistent with each other. 
The observational constraints clearly prefer a model in which star formation in low-mass halos 
changes behavior at a characteristic redshift $z_c\sim 2$. There is also 
a tentative evidence that this characteristic redshift depends on environments, 
becoming $z_c\sim 4$ in regions that eventually evolve into rich clusters 
of galaxies. However, given the uncertainties of the current observed CSMFs in the low-mass ends, this environmental dependence of $z_c$ needs to be confirmed with better data. 

We compare the predictions of a number of popular halo-based 
empirical models and two numerical simulations of galaxy formation.  
We find that the two numerical simulations fail to match the observational data one way or another. 
The empirical models by \citet{yang12} and \citet{moster13} fail to reproduce the faint-end upturn of the field SMFs from observations. 
The model by \citet{behroozi13} reproduces the faint-end upturn, but
it is a combined result of over-prediction for central galaxies 
and under-prediction for satellites at the faint-end. 
In contrast, the model by \citet{lu14,lu15} matches reasonably well the CSMFs in halos 
of different masses. The Lu et al. model predicts a much higher star formation  efficiency than the other models
for low-mass halos at redshifts higher than a characteristic redshift after which the star formation is suppressed. 

We use our constrained model to make predictions for a number of 
statistical properties of the galaxy population. These 
include the stellar mass functions of galaxies at high $z$, 
the stellar mass - halo mass relations at different redshifts, 
and the star formation and stellar mass assembly histories of galaxies 
in dark matter halos of different masses. A comparison of our model predictions 
with those of other empirical models shows that different models can make 
vastly different predictions for these properties, even though all 
of them are tuned to match the observed stellar mass functions of galaxies.
In particular, our constrained model predicts a much higher
{\it in situ} star formation rate at $z\geq2$ for present-day dwarf galaxies than the other models. As a result, such galaxies have about $40\%$ of their current-day stellar mass already in place by $z\sim2$.
Because of this boosted star formation in low-mass halos at high $z$, 
the role of accretion of stars from satellite galaxies, relative to 
{\it in situ} star formation, in the build up of massive galaxies 
is more important in our model than in the other models. 

One of the main predictions of our constrained model is the existence of 
a characteristic redshift that separates an active star formation phase
from a subdued star formation phase in low-mass halos. This change in star 
formation mode is likely related to the feedback processes that regulate 
star formation. As discussed in \citet{lu14,lu15}, energy feedback from stars 
and AGNs associated with active star formation and super-massive 
black hole accretion at high redshift may preheat the gas media around 
dark matter halos and suppress gas accretion and star formation at lower 
redshift \citep[][]{mo02, mo04}. Based on plausible assumptions 
about the star formation histories of the universe and the density of the 
intergalactic medium, the pre-heating is expected to occur around $z=2-3$, 
and the specific entropy of the preheated gas is $\sim 10{\rm KeV cm^2}$, 
which is important in affecting star formation in low-mass halos, because 
of their relatively shallow gravitational potential wells, but has no 
significant effects on halos with masses above $\sim 10^{12}\,{\rm M}_\odot$
\citep[e.g.][]{lu07}. This preheating may also explain why the cold gas mass function 
at $z\sim 0$ is shallow \citep{mo05}.  In such a scenario, the 
pre-heating is expected to occur earlier in regions occupied 
by present-day massive halos, because intensive star formation 
and AGN activity are expected to occur earlier in higher density  
regions where gravitational collapse is more accelerated. 
Our tentative finding of the positive dependence of the characteristic 
redshift on halo mass is in agreement with such an expectation, but 
better observational data are needed in order to examine such dependence in more detail.

\section*{Acknowledgments}
HJM acknowledges the support from NSF AST-1517528. BM and TWL acknowledge
support from NSF-1313302.

We thank Zhankui Lu for providing his codes for the star formation model implementation
and the merger tree generation, and Farhan Feroz for providing the source code of MULTINEST. 
We appreciate helpful discussions and comments from Sandy Faber, Frank van den Bosch, 
and Simon White, and many suggestions from Darren Croton, the referee, which improved
this paper. 

We acknowledge the Virgo Consortium for making the EAGLE simulation data available. 
The EAGLE simulations were performed using the DiRAC-2 facility at Durham, 
managed by the ICC, and the PRACE facility Curie based in France at TGCC, 
CEA, Bruy\`eresle-Ch\^atel. We are also grateful to all the involved groups for
making the Illustris simulation data available.


\label{lastpage}

\end{document}